\documentclass[%
 reprint,
superscriptaddress,
bibnotes,
 amsmath,amssymb,
 aps, prl, 
]{revtex4-2}

\usepackage{graphicx}
\usepackage{dcolumn}
\usepackage{bm}
\usepackage{braket}
\usepackage[colorlinks=true, allcolors=blue]{hyperref}
\usepackage{cleveref}
\usepackage{xeCJK}
\usepackage{svg}

\begin{document}

\title{Carrier Capture at Defects from Finite-Temperature Lattice Dynamics}
\author{Menglin Huang}
\email{menglinhuang@fudan.edu.cn}
\affiliation{College of Integrated Circuits and Micro-Nano Electronics, and Key Laboratory of Computational Physical Sciences (MOE), Fudan University, Shanghai 200433, China}
\author{Shanshan Wang}
\affiliation{College of Integrated Circuits and Micro-Nano Electronics, and Key Laboratory of Computational Physical Sciences (MOE), Fudan University, Shanghai 200433, China}
\affiliation{Department of Physics, Shanghai Normal University, Shanghai 200234, China}
\author{Shiyou Chen}
\email{chensy@fudan.edu.cn}
\affiliation{College of Integrated Circuits and Micro-Nano Electronics, and Key Laboratory of Computational Physical Sciences (MOE), Fudan University, Shanghai 200433, China}

\begin{abstract}
Defect-assisted carrier capture is commonly described within nonradiative multiphonon (NMP) theory using normal modes of the equilibrium defect structure. This description becomes inadequate when finite-temperature lattice fluctuations explore configurations that cannot be represented by a fixed normal-mode basis. Here, we present a trajectory-based method for calculating carrier capture rate from first principles, which allows lattice relaxation and electron--lattice coupling matrix element to be reconstructed directly from correlation functions of finite-temperature lattice dynamics. For hole capture at C$_\mathrm{N}$ in GaN within harmonic regime, the method reproduces static NMP results including mode mixing and agrees with experiment. For oxygen vacancy in SiO$_2$, by contrast, the thermally sampled potential energy surface is substantially softer than the zero-temperature normal-mode harmonic expansion, strongly modifying the lattice relaxation and electron--lattice coupling, and producing pronounced changes in both the capture coefficient and its temperature dependence. These results establish the finite-temperature configurational ensemble, rather than phonon occupations alone, as an essential ingredient of defect-assisted carrier capture.

\end{abstract}

\maketitle

Defect-assisted nonradiative carrier capture is a key process controlling carrier lifetimes and device performance in semiconductors \cite{Lang1975PRL,Guo2025PRA}. Its microscopic description is provided by nonradiative multiphonon (NMP) theory \cite{Huang1950,Kubo1955,Henry1977PRB,Huang1981}, in which carrier capture is accompanied by lattice relaxation and the rate is governed by vibrational overlap and off-diagonal electron--lattice coupling. Modern first-principles implementations generally describe this process using normal modes constructed around a zero-temperature equilibrium defect structure \cite{Shi2012PRL,Alkauskas2014PRB,Shi2015PRB,Nichols2025PRB,Zhou2025PRB}, with finite-temperature effects entering only through the thermal occupations of these modes \cite{Huang1981}.

This static normal-mode picture becomes inadequate when finite-temperature lattice dynamics explore configurations beyond the static region around a single equilibrium structure, as can occur for large structural relaxation \cite{Kim2019PRB}, soft lattices, dynamic disorder, or amorphous systems \cite{Gehrmann2019, Zacharias2023}. In such cases, the local curvatures and vibrational eigenvectors relevant to the carrier capture may vary along the trajectory. Previous extensions have introduced one-dimensional anharmonic potential-energy surfaces (PESs) \cite{Kim2019PRB, ZhangPRB2020,Kavanagh2021} or effective mode-frequency corrections \cite{Xiao2020}, but the lattice dynamics remain represented through predefined static coordinates. As long as lattice relaxation and electron--lattice coupling are formulated in terms of such predefined coordinates, extending the theory to general finite-temperature lattice dynamics remains conceptually and computationally challenging.

Here, we propose a trajectory-based method for calculating nonradiative carrier capture rate without normal-mode representation. Lattice relaxation and electron--lattice coupling matrix element, conventionally expressed through normal-mode-resolved quantities, are recast as correlation functions evaluated directly along finite-temperature molecular-dynamics (MD) trajectories. For hole capture at C$_\mathrm{N}$ in GaN, where the lattice dynamics remain harmonic, the method reproduces static NMP results including mode mixing and agrees with experiment. We then apply the method to oxygen vacancy in amorphous SiO$_2$, an example beyond harmonic approximation, and find the thermally sampled PES is substantially softer than that from the normal-mode analysis. This finite-temperature restructuring of lattice relaxation and electron--lattice coupling leads to pronounced changes in the capture coefficient and its temperature dependence, demonstrating that carrier capture can depend on the configurational ensemble rather than only on thermal phonon occupations.

To extend carrier-capture calculations to finite-temperature lattice trajectories, we start from the time-dependent formulation of the transition rate \cite{Kubo1955}. Using the coherent-state displacement operator to account for lattice relaxation and applying Kubo's cumulant expansion \cite{Kubo1962cumulant}, the transition rate within the static-coupling NMP formalism can be written as (see the Supplemental Material (SM) for the operator derivation \cite{SM} \nocite{Glauber1963, Cahill1969,huang2026DLTS,heyd2003hybrid,PBE,kresse1996efficiency,Batzner2022,Kavanagh-model,zhengqijing})
\begin{equation}\label{second_rate}
r=\frac{1}{\hbar^2}\int_{-\infty}^{\infty}dt\,e^{i\Delta Et/\hbar}G(t)\left[C_{VV}(t)+K_{\mathrm{HT}}(t)\right].
\end{equation}
Here, $\Delta E$ is the electronic transition energy, $G(t)$ describes lattice relaxation, $C_{VV}(t)$ describes fluctuations of the off-diagonal electron--lattice coupling matrix element $V_{if}=\langle\psi_i|H_{\mathrm{eL}}|\psi_f\rangle$, and $K_{\mathrm{HT}}(t)$ describes their cross correlation. In the conventional normal-mode representation, these three quantities are determined by the mode-resolved Huang--Rhys factors $S_k$, electron--lattice coupling strengths $\partial V_{if}/\partial Q_k$, and signed cross weights $\Delta Q_k\,\partial V_{if}/\partial Q_k$, respectively \cite{Zhou2025PRB}. Their explicit normal-mode expressions and spectral representations are given in the SM \cite{SM}. We next show that the same three ingredients can be reconstructed directly from finite-temperature MD trajectories without an explicit normal-mode decomposition.

For lattice relaxation term $G(t)$, the MD-based spectral function has been calculated in quantum chemistry for luminescence applications \cite{Valleau2012,Linderlv2025}. Following the similar idea, here we generate a trajectory $\mathbf{R}(t)$ on the initial-state PES and evaluate the transition energy between the initial and final states at the same instantaneous configuration. We define the fluctuation of the electronic transition energy as $\delta E(t)=U_i[\mathbf{R}(t)]-U_f[\mathbf{R}(t)]-\langle U_i-U_f\rangle$ and its classical correlation function as $C_{EE}^{\mathrm{cl}}(t)=\langle\delta E(0)\delta E(t)\rangle$. In the quantum harmonic picture \cite{Huang1981}, its frequency spectrum is directly related to the Huang--Rhys spectral function, yielding \cite{SM}
\begin{equation}\label{S_MD}
S_{\mathrm{MD}}(\omega)=\frac{1}{k_BT\hbar\omega}\left[\frac{1}{2\pi}\int_{-\infty}^{\infty}dt\,e^{i\omega t}C_{EE}^{\mathrm{cl}}(t)\right].
\end{equation}
Since $G(t)$ is fully determined by the Huang--Rhys spectral function \cite{Alkauskas2014NJP,Jin2021PRM}, Eq.~(\ref{S_MD}) allows the lattice-relaxation contribution to be reconstructed directly from the MD trajectory.

The electron--lattice coupling matrix element can likewise be evaluated without normal modes. Using the completeness of the normal-mode eigenvectors together with the chain rule, its mode expansion is exactly transformed into Cartesian coordinates \cite{SM},
\begin{equation}\label{Vif_cartesian}
V_{if}(t)=\sum_\alpha\frac{\partial V_{if}}{\partial\mathbf{R}_\alpha}\cdot\Delta\mathbf{R}_\alpha(t),
\end{equation}
where $\Delta\mathbf{R}_\alpha(t)=\mathbf{R}_\alpha(t)-\mathbf{R}_\alpha^{\mathrm{ini}}$, $\alpha$ is the atom order in the supercell. The classical trajectory therefore directly provides $C_{VV}^{\mathrm{cl}}(t)=\langle V_{if}(0)V_{if}(-t)\rangle$. Because a classical correlation function does not reproduce the quantum asymmetry between multiphonon emission and absorption \cite{Huang1950}, we impose the detailed-balance correction in the frequency domain \cite{SM}. Defining $\widetilde{C}_{VV}^{\mathrm{cl}}(\omega)=(2\pi)^{-1}\int_{-\infty}^{\infty}d\tau\,e^{i\omega\tau}C_{VV}^{\mathrm{cl}}(\tau)$, the quantum-corrected correlation function becomes \cite{SM}
\begin{equation}\label{C_VV_q}
\begin{aligned}
C_{VV}^{\mathrm q}(t)=&\int_{-\infty}^{\infty}d\omega\,e^{-i\omega t}f_{VV}(\omega)\widetilde{C}_{VV}^{\mathrm{cl}}(\omega),\\
f_{VV}(\omega)=&\frac{\hbar\omega}{k_BT\left(1-e^{-\hbar\omega/k_BT}\right)}.
\end{aligned}
\end{equation}

Finally, the cross term $K_{\mathrm{HT}}(t)$ is governed by the signed mode weight $\Delta Q_k\,\partial V_{if}/\partial Q_k$. This information can be extracted from a velocity cross correlation along the MD trajectory. With $\Delta\mathbf{R}_\alpha=\mathbf{R}_\alpha^{\mathrm{fin}}-\mathbf{R}_\alpha^{\mathrm{ini}}$, we define $C_{\mathrm{HT}}(t)=\left\langle\dot V_{if}(0)\sum_\alpha m_\alpha\Delta\mathbf{R}_\alpha\cdot\dot{\mathbf{R}}_\alpha(t)\right\rangle$. In the quantum picture, the momentum correlation function gives $C_{\mathrm{HT}}(t)=k_BT\int_0^\infty d\omega\,J_{\mathrm{HT}}(\omega)\cos(\omega t)$, and hence \cite{SM}
\begin{equation}\label{J_HT_final}
J_{\mathrm{HT}}(\omega)=\frac{2}{\pi k_BT}\int_0^\infty dt\,C_{\mathrm{HT}}(t)\cos(\omega t).
\end{equation}
The resulting $J_{\mathrm{HT}}(\omega)$ determines $K_{\mathrm{HT}}(t)$ through the corresponding spectral representation given in the SM \cite{SM}.

\begin{figure*}[htbp]
	\centering
	\includegraphics[width=0.9\textwidth]{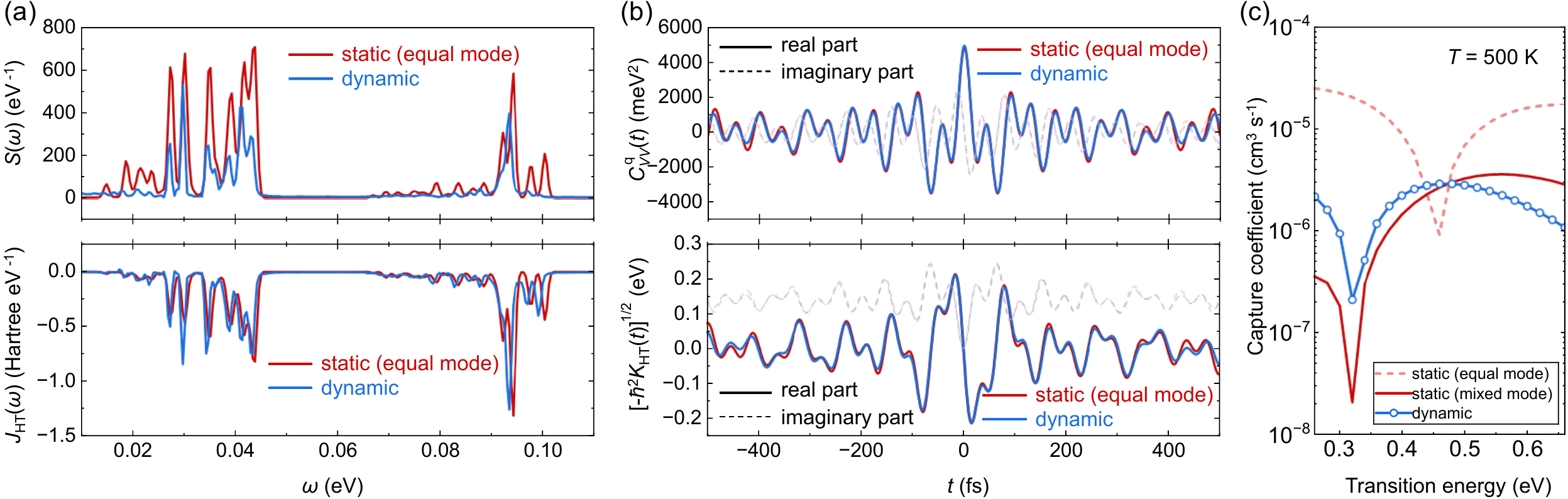}
	\caption{(a) Huang--Rhys spectral function $S(\omega)$ and cross spectrum of lattice relaxation with electron--lattice coupling $J_{\mathrm{HT}}(\omega)$ obtained at 500 K using the static initial-state equal-mode method and the dynamical method. (b) Real and imaginary parts of the quantum-corrected electron--lattice coupling correlation function $C_{VV}^{\mathrm q}(t)$ and the cross-coupling term $[-\hbar^2K_{\mathrm{HT}}(t)]^{1/2}$. (c) Hole-capture coefficient as a function of the electronic transition energy obtained using the static equal-mode approximation, static mode-mixing, and dynamical approach.}
	\label{fig1}
\end{figure*}

For general finite-temperature trajectories, we regard the spectral quantities reconstructed from MD, with quantum corrections introduced where needed, as effective spectral functions. Finite-temperature structural fluctuations and anharmonicity thus enter through the configurations sampled by the trajectory itself. Equations show above provide the carrier capture rate directly from finite-temperature lattice dynamics.

To establish the connection between the present approach and conventional static methods, we first consider hole capture at C$_\mathrm{N}$ in GaN. The lattice relaxation in this system remains within the harmonic regime \cite{Alkauskas2014PRB,Zhou2025PRB}, making it a suitable benchmark for the trajectory formulation. Nevertheless, the normal modes and frequencies of the initial and final states differ appreciably, such that the commonly used equal-mode approximation is insufficient \cite{Zhou2025PRB}. Our previous static NMP treatment accounts for this renormalization through the Duschinsky transformation and quadratic electron--lattice cross terms, referred to below as the mode-mixing method \cite{Zhou2025PRB}. We therefore compare the dynamical results with both the equal-mode and mode-mixing static results.

Fig.~\ref{fig1}(a) compares the Huang--Rhys spectrum $S(\omega)$ and the cross spectrum $J_{\mathrm{HT}}(\omega)$ obtained from the static equal-mode representation (Huang-Rhys factor can only be defined under the equal-mode approximation) and from the MD trajectory. The dominant spectral structures are similar, although the dynamical $S(\omega)$ has a noticeably smaller overall weight, whereas the signed $J_{\mathrm{HT}}(\omega)$ agrees well in peak positions and relative positive--negative structure. Consistently, Fig.~\ref{fig1}(b) shows close agreement between the static and dynamical $C_{VV}^{\mathrm q}(t)$ and $[-\hbar^2K_{\mathrm{HT}}(t)]^{1/2}$. The remaining difference therefore originates mainly from the lattice-relaxation spectrum. Despite this difference, Fig.~\ref{fig1}(c) shows that the dynamical capture coefficients closely reproduce the static mode-mixing results, agreeing within one order of magnitude and yielding the same position of the central minimum. This minimum corresponds to the transition energy at which the effective capture barrier vanishes, showing that our dynamical method correctly describe the multidimensional PESs during carrier capture.

\begin{figure}[htbp]
	\centering
	\includegraphics[width=0.48\textwidth]{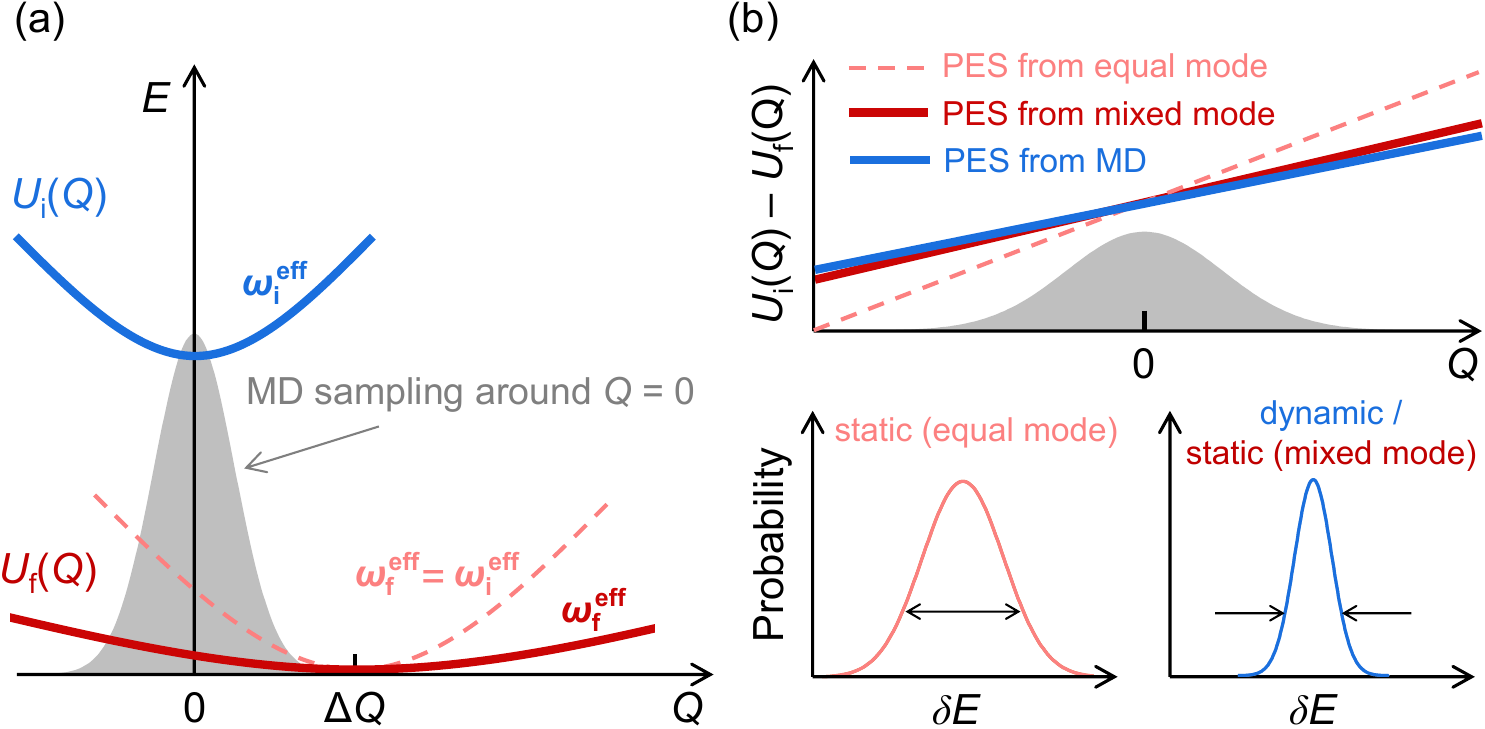}
	\caption{(a) Schematic potential-energy surfaces of the initial and final states projected onto the effective lattice-relaxation coordinate. The gray distribution indicates finite-temperature MD sampling around $Q=0$ on the initial-state potential-energy surface, while the dashed curve represents the final-state potential-energy surface under the equal-mode approximation, in which the final-state frequency is replaced by the initial-state frequency. (b) Upper panel: initial-to-final energy difference $U_i-U_f$ predicted by different descriptions of the potential-energy surfaces within the configurational window sampled from the initial state. Lower panel: corresponding schematic distributions of the transition-energy fluctuations.}
	\label{fig2}
\end{figure}

The origin of this agreement is illustrated in Fig.~\ref{fig2}. Projecting the initial- and final-state PESs onto the effective lattice-relaxation coordinate gives $\omega_i^{\mathrm{eff}}=43.7$ meV and $\omega_f^{\mathrm{eff}}=38.9$ meV, respectively (see the SM \cite{SM}), indicating that the initial-state PES is steeper than the final-state PES. Because the MD trajectory samples the initial-state surface around $Q=0$, the relevant quantity is the energy difference $U_i-U_f$ evaluated at the same instantaneous configurations. The dynamical approach therefore retains the actual curvature difference between the two PESs without explicitly constructing the final-state normal modes or the Duschinsky matrix. By contrast, the equal-mode approximation forces the final-state curvature to equal that of the initial state, leading to artificially enhanced fluctuations of $U_i-U_f$ within the sampled configurational window [Fig.~\ref{fig2}(a,b)]. Since $C_{EE}(0)=\mathrm{Var}_i[U_i-U_f]$, these larger transition-energy fluctuations translate directly into the larger $S(\omega)$ obtained from the equal-mode approximation.

Thus, although the trajectory formulation does not explicitly construct normal modes and Duschinsky matrix, the underlying differences between the initial- and final-state multidimensional PESs enter naturally through the instantaneous transition-energy fluctuations. The resulting capture coefficients reproduce the static mode-mixing calculation and are consistent with experiment \cite{SM, Reshchikov2014, Reshchikov2021}.

\begin{figure}[htbp]
	\centering
	\includegraphics[width=0.48\textwidth]{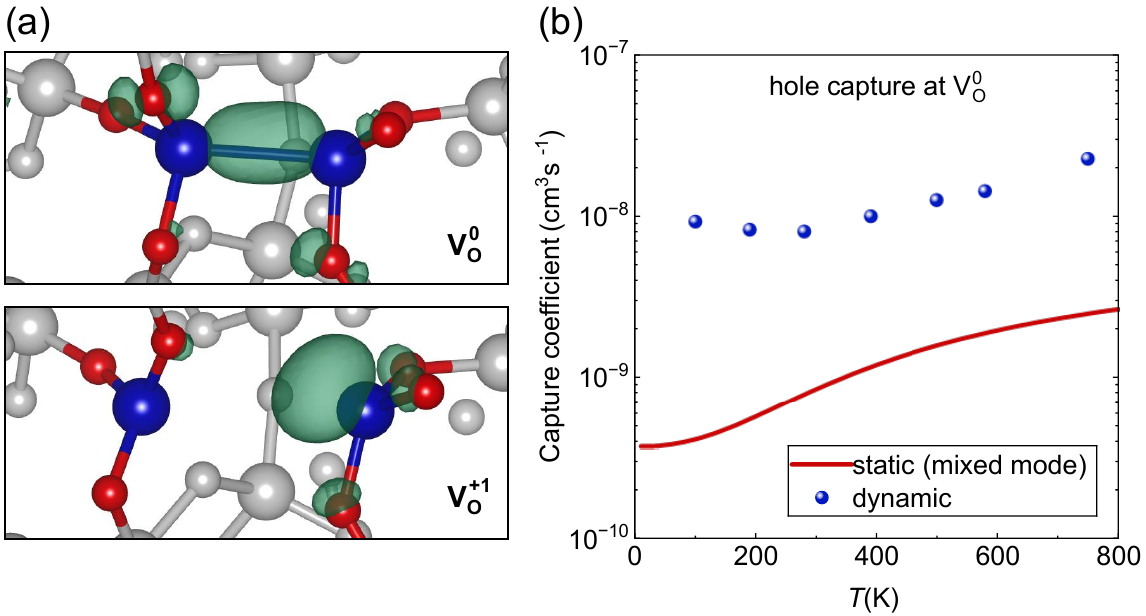}
	\caption{(a) Defect structures and local charge densities of the neutral and $+1$ charge states of $V_\mathrm{O}$ in SiO$_2$. (b) Temperature dependence of the $V_\mathrm{O}$ hole-capture coefficient calculated using the conventional static mode-mixing method and the dynamical method.}
	\label{fig3}
\end{figure}

We next turn to hole capture by an oxygen vacancy in amorphous SiO$_2$, where finite-temperature lattice dynamics deviate substantially from the static normal-mode picture. Hole capture converts $V_{\mathrm O}^{0}$ into $V_{\mathrm O}^{+1}$ and produces pronounced changes in the defect wavefunction and local structure while preserving the Si--Si dimer configuration \cite{Guo2025PRA}. As shown in Fig.~\ref{fig3}(b), the dynamical approach yields substantially larger capture coefficients than the static mode-mixing calculation over the entire temperature range, with a clear discrepancy already at $100~\mathrm{K}$. The two methods also exhibit distinct temperature dependences, indicating that the difference cannot be described solely by changing the thermal occupations of a fixed set of zero-temperature normal modes.

\begin{figure}[htbp]
	\centering
	\includegraphics[width=0.5\textwidth]{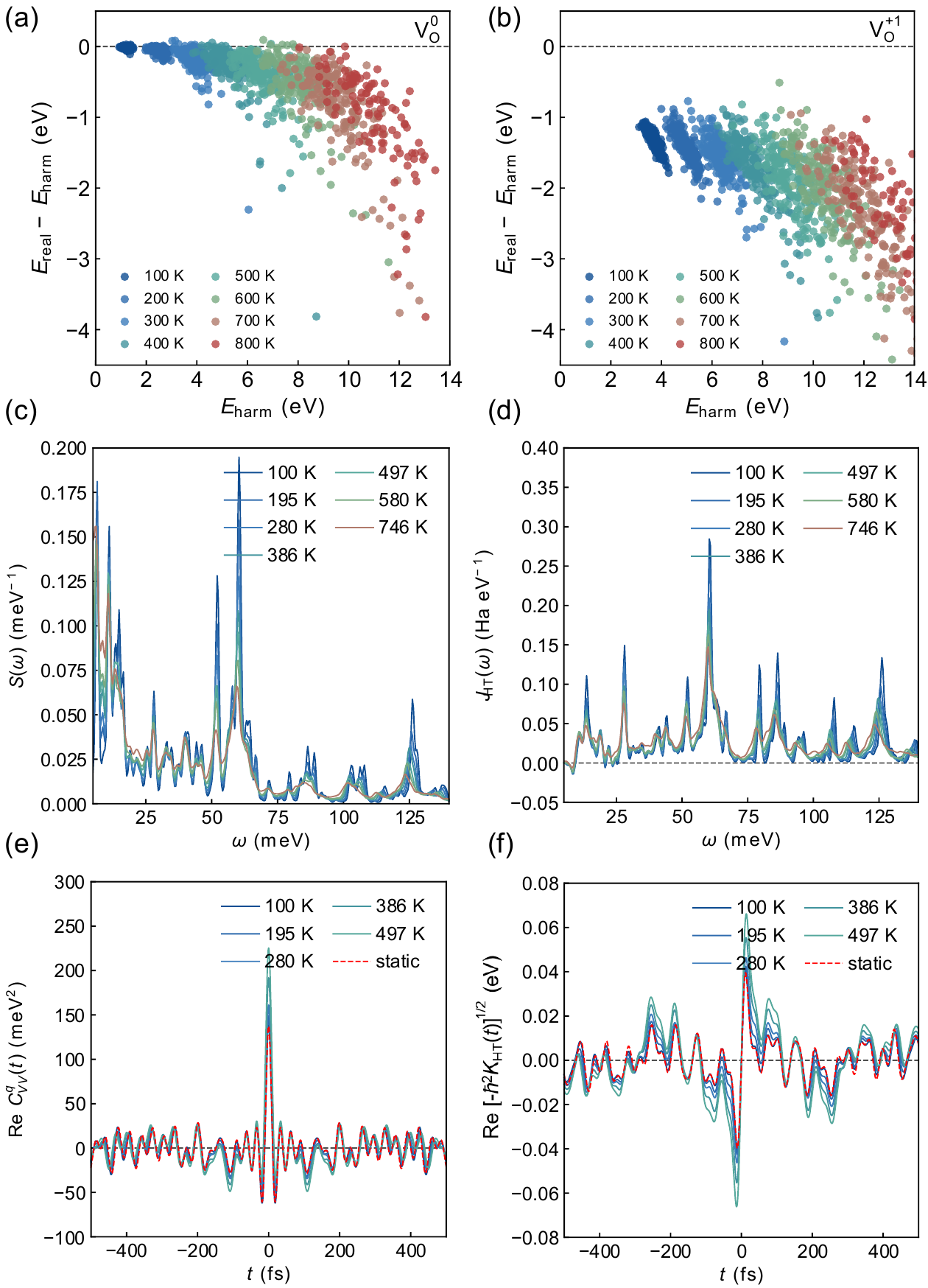}
	\caption{(a,b) Difference between the explicitly calculated and harmonic energies, $E_{\mathrm{real}}-E_{\mathrm{harm}}$, for the $V_{\mathrm O}^{0}$ and $V_{\mathrm O}^{+1}$ potential-energy surfaces, respectively, evaluated along finite-temperature trajectories of $V_{\mathrm O}^{0}$. (c,d) Huang--Rhys spectral function $S(\omega)$ and cross spectrum $J_{\mathrm{HT}}(\omega)$, respectively, obtained from the dynamical approach at different temperatures. (e,f) Real parts of the quantum-corrected electron--lattice coupling correlation function $C_{VV}^{\mathrm q}(t)$ and the cross-coupling term $[-\hbar^2K_{\mathrm{HT}}(t)]^{1/2}$, respectively, at different temperatures.}
	\label{fig4}
\end{figure}

To test the static harmonic representation directly, we compare the explicitly calculated total energy $E_{\mathrm{real}}$ of configurations sampled along the $V_{\mathrm O}^{0}$ trajectories with the quadratic energy $E_{\mathrm{harm}}$ predicted from the zero-temperature Hessian on both the $V_{\mathrm O}^{0}$ and $V_{\mathrm O}^{+1}$ PESs. If the sampled configurations were accurately described by the normal-mode expansion, $E_{\mathrm{real}}\simeq E_{\mathrm{harm}}$ would hold. Instead, Figs.~\ref{fig4}(a) and \ref{fig4}(b) show that $E_{\mathrm{real}}-E_{\mathrm{harm}}<0$ systematically for both charge states, with the deviation increasing as the trajectory explores a broader configurational region at higher temperatures. Importantly, the deviation is already evident at $100~\mathrm{K}$, particularly for $V_{\mathrm O}^{+1}$. Thus, the configurations relevant to carrier capture experience a substantially softer PES than that represented by the zero-temperature harmonic normal-mode expansion.

This softening is directly reflected in the lattice-relaxation spectrum. Fig.~\ref{fig4}(c) shows $S_{\mathrm{MD}}(\omega)$ obtained from the transition-energy correlation function, whereas the corresponding static spectrum, provided in the SM \cite{SM}, has a substantially larger spectral weight. The quantum-corrected $C_{EE}^{\mathrm q}(0)$ reconstructed from $S_{\mathrm{MD}}(\omega)$ gives a PL linewidth of $\mathrm{FWHM}=2\sqrt{2\ln2\,C_{EE}^{\mathrm q}(0)}=0.31$ eV, within the experimentally reported range for oxygen-vacancy-related centers in SiO$_2$ \cite{Nishikawa1996JAP,Sakurai2004JNS}. By contrast, the static and dynamical $J_{\mathrm{HT}}(\omega)$ in Fig.~\ref{fig4}(d) differ by less than one order of magnitude. This weaker sensitivity is consistent with $S_k\propto\Delta Q_k^2$ but $J_{\mathrm{HT},k}\propto\Delta Q_k$, making the Huang--Rhys spectrum more sensitive to changes in the lattice-relaxation amplitudes.

The temperature dependence extends beyond $S_{\mathrm{MD}}(\omega)$. Both $C_{VV}^{\mathrm q}(t)$ and the cross term constructed from $J_{\mathrm{HT}}(\omega)$ vary appreciably with temperature [Figs.~\ref{fig4}(e) and \ref{fig4}(f)], showing that the effective electron--lattice coupling and its correlation with lattice relaxation also evolve with the sampled configurations. Temperature therefore changes not only the occupations of vibrational excitations but also the configurational ensemble entering the transition. In contrast, the analysis for C$_\mathrm{N}$ in GaN shows no comparable deviation (see the SM \cite{SM}). The comparison for V$_\text{O}$ here is mainly for demonstrations; the actual hole capture behavior in amorphous system requires statistical average of carrier capture on different oxygen sites, which is beyond the scope of the present study.

In summary, we have proposed a finite-temperature trajectory-based method for calculating nonradiative carrier capture rate from first principles. The method reconstructs lattice relaxation and electron--lattice coupling matrix element directly from molecular-dynamics correlation functions without requiring a fixed normal-mode basis. We apply the method to two examples, which reproduces static NMP results including mode mixing for C$_\mathrm{N}$ in GaN, while reveals qualitatively different carrier capture behavior for an oxygen vacancy in amorphous SiO$_2$, where the configurations sampled at finite temperature experience a substantially softer PES than predicted by the zero-temperature harmonic normal-mode expansion. These results imply that the configurational ensemble, rather than phonon thermal occupations alone, is an indispensable part of defect-assisted carrier capture at finite temperatures. The method presented here is general for treating carrier capture in defect systems with large lattice relaxation, dynamic disorder, and soft lattices.

The data that support the findings of this article will be openly available once the manuscript has been accepted for publication.

This work was supported by National Natural Science Foundation of China (12334005, 12188101 and 12404089), National Key Research and Development Program of China (2024YFB4205002), Science and Technology Commission of Shanghai Municipality (24JD1400600).

M. Huang and S. Wang contributed equally to this work.

\nocite{*}

\bibliography{aps}

\end{document}